\documentclass[12pt]{article}
\usepackage{amssymb}
\usepackage{amsmath}
\usepackage{hyperref}
\begin{document}
\title{\bf Holographic Aspects of Non-minimal $R_{\mu \alpha \nu \beta } F^{(a)\mu \alpha } F^{(a)\nu \beta } $ AdS Black Brane}
\author{ %Shahrokh Parvizi$^1$\thanks{Corresponding author: Email:parvizi@modares.ac.ir}  \hspace{2mm} and
      Mehdi Sadeghi\thanks{Email:  mehdi.sadeghi@abru.ac.ir}\hspace{2mm}\\
	%	{\small {\em $^1$Department of Physics, School of Sciences,}}\\
     %   {\small {\em Tarbiat Modares University, P.O.Box 14155-4838, Tehran, Iran }}\\
{\small {\em Department of Physics, School of Sciences,}}\\
        {\small {\em Ayatollah Boroujerdi University, Boroujerd, Iran}}
       }
\date{\today}
\maketitle

\abstract{ In this paper, we study the holographic dual to an asymptotically anti-de Sitter black brane in an Einstein-Yang-Mills model with a non-minimal coupling between the Riemann and Yang-Mills fields. First, we construct a planar black hole solution of this model up to the first order of the non-minimal coupling of the Yang-Mills field with the Riemann-Christoffel tensor, denoted as $q_2$. Then, we calculate the color non-abelian direct current (DC) conductivity and the ratio of shear viscosity to entropy density for this solution. Our result for the shear viscosity $\eta$ to entropy density $s$ ratio saturates the Kovtun, Son, and Starinets (KSS) bound, which is proportional to $\frac{1}{4 \pi}$. However, our result for the conductivity is new up to the first order of $q_2$.}\\

%\pacs{11.10.Jj, 11.10.Wx, 11.15.Pg, 11.25.Tq}

\section{Introduction} \label{intro}
 Hydrodynamics is an effective theory for a quantum many-body system in the low-energy limit. Hydrodynamics equations are written in terms of Neother charges. Conservation laws for the stress-energy tensor and internal currents are the equations of hydrodynamics.
 \begin{align}
 & \nabla _{\mu } J^{\mu } =0\,\,\,\,\,,\,\,\nabla _{\mu } T^{\mu \nu} =0,\\
 & J^{\mu } =n \, u^{\mu }-\sigma T P^{\mu \nu }\partial_{\nu}(\frac{\mu}{T})\nonumber\\
 & T^{\mu \nu } =(\rho +p)u^{\mu } u^{\nu } +pg^{\mu \nu } -\sigma ^{\mu \nu },\nonumber\\
 &\sigma ^{\mu \nu } = {P^{\mu \alpha } P^{\nu \beta } } [\eta(\nabla _{\alpha } u_{\beta } +\nabla _{\beta } u_{\alpha })+ (\zeta-\frac{2}{3}\eta) g_{\alpha \beta } \nabla .u]\nonumber\\& P^{\mu \nu }=g^{\mu \nu}+u^{\mu}u^{\nu}, \nonumber
 \end{align}
 \indent where $n$, $\sigma$, $\eta$, $\zeta $, $\sigma ^{\mu \nu }$ and $P^{\mu \nu }$ are charge density, charge conductivity, shear viscosity, bulk viscosity, shear tensor, and projection operator, respectively \cite{Kovtun2012, Bhattacharyya,Rangamani,J.Bhattacharya2014}.\\

The properties of a fluid are determined by its transport coefficients. Therefor, we are interested in calculating the color non-abelian DC conductivity\cite{Litim:1999id} and the ratio of shear viscosity to entropy density as transport coefficients in hydrodynamic equations up to the first order of derivative expansion.\\

The Anti-de Sitter/Conformal Field Theory duality, also known as the AdS/CFT correspondence \cite{Maldacena,Aharony}, is an important conjecture in string theory. It states that a string theory on $AdS_5$ is equivalent to a $\mathcal{N} = 4$ super-Yang-Mills theory in 4-dimensional spacetime. This duality is generalized to the gauge/gravity correspondence. The fluid-gravity duality \cite{Bhattacharyya} results in the AdS/CFT duality in the low energy limit. This conjecture opens up new possibilities for understanding a strongly interacting gauge theory through the study of a dual weakly interacting gravitational theory.\\

Theories that couple the gravitational field to other fields using cross terms containing the curvature tensor are called non-minimal theories\cite{Balakin:2015gpq}. The non-minimal term between the Ricci scalar and the Yang-Mills field was investigated in \cite{Sadeghi:2023hxd}. In this paper, we study the non-minimal term between Riemann tensor and the Yang-Mills field. They are introduced as alternative theories of gravity, and they have five classes. Exact solutions of stars \cite{Horndeski:1978ca,Mueller-Hoissen:1988cpx}, wormholes \cite{Balakin:2007xq,Balakin:2010ar}, and regular magnetic BHs \cite{Balakin:2006gv} with Wu-Yang anstaz \cite{Wu2007,Shnir:2005xx} are constructed by these non-minimal theories. Power-law inflation can be realized due to the non-minimal gravitational coupling of the electromagnetic field, and large-scale magnetic fields can be generated due to the breaking of the conformal invariance of the electromagnetic field through its non-minimal gravitational coupling\cite{Bamba:2008ja}.\\

We use the Green-Kubo \cite{Son},\cite{Policastro2001} formula to calculate conductivity as follows,
\begin{equation} \label{kubo}
\sigma^{ij} (k_{\mu})=-\mathop{\lim }\limits_{\omega \to 0} \frac{1}{\omega } \Im G^{ij}(k_{\mu})
\end{equation} 
where $a,b$ and $i,j$ indices referring to the $SL(2,C)$ group symmetry. There is a universal bound for the conductivity, $\sigma > 1$ where $\sigma$ is measured in units of $\frac{e^2}{\hbar}$ and $e$ represents the charge carried by the gauge field \cite{Grozdanov:2015qia}. This bound is violated in various theories, including massive gravity \cite{Grozdanov:2015qia}, models with background fields \cite{Donos:2014cya}, non-abelian Born-Infeld theory\cite{Sadeghi:2021qou}, non-abelian logarithmic type of Born-Infeld theory\cite{Sadeghi:2022mog} and a system with disorder\cite{Baggioli:2016oqk}. We want to investigate whether this bound is preserved for our model or not.\\

$Dp$-branes are electrically charged in string theory. In type-IIA superstring theory, $p$ is even and couples to $q$-forms with odd $q$. But, $p$ is odd in type-$IIB$ superstring theory, and it couples to the $q$-forms with even $q$.\\

The NS5-brane is a five-dimensional $p$-brane coupled to the Kalb-Ramon field (B-field) magnetically, but fundamental strings are coupled to the Kalb-Ramon field electrically. They are BPS states in which the mass is equal to the charge, $M=Q$. Branes are stable in string theory, so the study of these branes is important. We mention that the black brane is the D-brane with one (or two) event horizon(s).\\

Electric black branes are useful for studying systems with a chemical potential in the field theory dual. Fluid with a chemical potential is an example of this kind of system.

Our motivation in this paper is to study the characteristics of the dual field theory by calculating transport coefficients such as conductivity and $\frac{\eta}{s}$.
%--------------------------------------------------------------------------
\section{   Non-minimal $R_{\mu \alpha \nu \beta } F^{(a)\mu \alpha } F^{(a)\nu \beta }$ AdS Black Brane Solution}
\label{sec2}
\indent Our non-minimal model is described by the bulk action as follows,
\begin{eqnarray}\label{action}
S=\int d^{4}  x\sqrt{-g} \bigg[\frac{1}{\kappa }(R-2\Lambda )-\frac{q_1}{4}F^{(a)}_{\mu \alpha }F^{ (a)\mu \alpha  }+q_2R_{\mu \alpha \nu \beta } F^{(a)\mu \alpha } F^{(a)\nu \beta } \bigg],
\end{eqnarray}
where $R$ is the Ricci scalar, $\Lambda=-\frac{3}{L^2}$ the cosmological constant, $L$ denotes the AdS radius, and $\mathcal{F}= F^{(a)}_{\mu \alpha }F^{ (a)\mu \alpha }$ represents the Yang-Mills invariant.
$F^{(a)}_{\mu \nu } $ is the Yang-Mills field strength tensor, 
\begin{align} \label{YM}
F^{(a)}_{\mu \nu } =\partial _{\mu } A^{(a)}_{\nu } -\partial _{\nu } A^{(a)}_{\mu } -i[A^{(a)}_{\mu }, A^{(b)}_{\nu }],
\end{align}
in which the gauge coupling constant is 1, $A^{(a)}_{\nu }$'s are the Yang-Mills potentials.
$R_{\mu \alpha \nu \beta }$ is the Riemann curvature tensor or Riemann-Christoffel tensor, where $q_1$ and $q_2$ are in generally arbitrary. $q_2$ is the phenomenological parameters describing the non-minimal coupling of the Yang-Mills field with the Riemann-Christoffel tensor\cite{Balakin:2005fu}.\\
We want to find the black brane solution for our non-minimal model. Therefore, the corresponding ansatz for the spacetime metric in the line element form is as,
\begin{equation}\label{metric}
ds^{2} =-f(z)e^{-2H(z)}dt^{2} +\frac{dz^{2} }{f(z)} +\frac{z^2}{L^2}(-dt^{2}+dx^2+dy^2).
\end{equation}
We consider the electic black brane in our model so the ansatz of the gauge potential is as,
\begin{equation}\label{background}
{\bf{A}} =\frac{1}{\sqrt{2}}h(z)dt\begin{pmatrix}1 & 0 \\ 0 & -1\end{pmatrix}.
\end{equation}
The gauge field has a gauge group $\sigma_3= \begin{pmatrix}1 & 0 \\ 0 & -1\end{pmatrix}$, which is expressed in terms of the diagonal generator of the $SL(2,C)$.\\

Varying the action (\ref{action}) with respect to the metric $g_{\mu \nu } $ yields the following Einstein's equation,
\begin{equation}\label{EOM1}
R_{\mu \nu }-  \tfrac{1}{2} g_{\mu \nu } R + \Lambda g_{\mu \nu }=\kappa T^{\text{(eff)}}_{\mu \nu }
\end{equation}
where,
\begin{equation}
T^{\text{(eff)}}_{\mu \nu }=q_1 T^{\text{(YM)}}_{\mu \nu } + q_2T^{(I)}_{\mu \nu }
\end{equation}
\begin{equation}
T^{\text{(YM)}}_{\mu \nu }= - \tfrac{1}{8}g_{\mu \nu } F^{(a)}_{\alpha \beta } F^{(a)\alpha \beta }+\tfrac{1}{2}F^{(a)}_{\mu }{}^{\alpha } F^{(a)}_{\nu \alpha },
\end{equation}
\begin{eqnarray}
&& -T^{(a)(I)}_{\mu \nu }=- \tfrac{1}{2} F^{(a)\alpha  \beta  } F^{(a)\gamma  \lambda  } g_{\mu  \nu  \
} R_{\alpha  \beta  \gamma  \lambda  } + \tfrac{3}{2} \
F^{(a)\beta  \gamma  } F_{\nu  }{}^{(a)\alpha  } R_{\mu  \alpha  \
	\beta  \gamma  } +\tfrac{3}{2} F^{(a) \beta  \gamma  } F^{(a)}_{\mu  \
}{}^{\alpha  } R_{\nu  \alpha  \beta  \gamma  } \nonumber \\ 
&&+ F^{(a)}_{\nu  \
}{}^{\alpha  } \nabla _{\alpha  }\nabla _{\beta  }F^{(a)}_{\mu  }{}^{\beta  \
}  + F^{(a)}_{\mu  }{}^{\alpha  } \nabla _{\alpha  }\nabla _{\beta  }F^{(a)}_{\nu  \
}{}^{\beta  } + 2 \nabla _{\alpha  }F^{(a)}_{\mu  }{}^{\alpha  } \nabla \
_{\beta  }F^{(a)}_{\nu  }{}^{\beta  } \nonumber \\ 
&& + F^{(a)}_{\nu  }{}^{\alpha  } \nabla \
_{\beta  }\nabla _{\alpha  }F^{(a)}_{\mu  }{}^{\beta  } + F^{(a)}_{\mu  \
}{}^{\alpha  } \nabla _{\beta  }\nabla _{\alpha  }F^{(a)}_{\nu  }{}^{\beta  \
} + 2 \nabla _{\alpha  }F^{(a)}_{\nu  \beta  } \nabla ^{\beta  }F^{(a)}_{\mu  \
}{}^{\alpha  }.
\end{eqnarray}
The Yang-Mills' equation is found by the varaition of the action (\ref{action}) with respect to the gauge field $A^{(a)}_{\mu}$,
 \begin{eqnarray}\label{EOM-YM}
\nabla_{\nu}\bigg(-\frac{1}{2}q_1 F^{(a)\mu \nu}+2 q_2 R^{\mu \nu \alpha \beta } F^{(a)}_{\alpha \beta} \bigg)=0.
\end{eqnarray}
The $(t,t)$ component of Einstein's field equations is found by using Eq.(\ref{EOM1}) as follows,
\begin{eqnarray}
&& -12 e^{-2 H}  \Big(f + \Lambda  z^2 + z f'\Big) \nonumber\\&& + z h' \Bigg( h' \
(-3 q_1 z + 8 q_2  f' (3 + 4 z \
H')  - 8 q_2 z f''
) + 8 q_2 z f' h''\Bigg)\nonumber\\&& + 16 q_2 \
f \Bigg(r^2 h''^2 + \
h'^2 (1 + z (H' (4 + z H') \nonumber\\&&+ 2 z H''
)) + z h' ((5 + 4 z H') \
h'' + z h''^{3}(z))\Bigg)=0,
\end{eqnarray}
for $q_2=0$ we have,
\begin{equation}\label{ttq2}
12 e^{-2 H(z)}  \Big(f + \Lambda  z^2 + z f'\Big)+3q_1 z^2 h'^2=0.
\end{equation}
The $(z,z)$ component of Einstein's equations is as follows,
\begin{eqnarray}
&& 4 f + 4 \Lambda  z^2 + 4 z f' - 8 \
f z H' + e^{2 H} z h' \Bigg(-8 q_2 z (f'\nonumber\\&&  -2 \
f H') \
h'' + h' (q_1 z  -8 f' (q_2 + 4 \
q_2 z H') + 8 q_2 (z \
f'' \nonumber\\&&+ 2 f (H' + 2 z (H'^2 - z \
H'')))\Bigg)=0,
\end{eqnarray}
for $q_2=0$ we have,
\begin{equation}\label{rrq2}
4 f+4 \Lambda  z^2+4 z f'+q_1 z^2 e^{2 H} h'^2-8 z f H'=0.
\end{equation}
By comparison between eq.(\ref{ttq2}) and eq.(\ref{rrq2}), we conclude,
\begin{equation}
 8 r f H'=0.
\end{equation} 
It means that $H(z)$ is a constant for $q_2=0$.\\
The $(x,x)$ component of eq.(\ref{EOM1}) is as follows,
\begin{eqnarray}\label{xx}
&&-r e^{2 H} h'^2 \left(4 q_2 \left(-3 f' H'+f''+2 f
\left(H'^2-H''\right)\right)+q_1\right)+f' \left(4-6 z H'\right)\nonumber\\&&+2 z \left(f''+2 \Lambda \right)+4 f \left(H'
\left(z H'-1\right)-z H''\right)=0.
\end{eqnarray}
For $q_2=0$ we have,
\begin{equation}
f' \left(4-6 z H'\right)+2 z \left(f''+2 \Lambda \right)+4 f \left(H' \left(z H'-1\right)-z H''\right)-q_1 z e^{2 H} h'^2=0.
\end{equation}
The $(y,y)$ component of gravitational field equations is the same as Eq.(\ref{xx}).\\
We solve the Yang-Mills' equations (\ref{EOM-YM}) and find $h(z)$ as below,
\begin{equation}
h(z)= C_1 \int_{z_h}
^{z}\frac{e^{-H(u)}
	}{\mathcal{B}(u)}du,
\end{equation}
where,
\begin{equation}
\mathcal{B}= u \Big(-4 q_2 f' \left(3 u H'+2\right)+4 q_2 u f''+3 q_1 u \Big)+8 q_2 f \Big(u \left(u H'^2+H'-u
H''\right)+1\Big).
\end{equation}
There is no exact solution to the field equations, so we attempt to find the solution to the first order of $q_2$. Therefore, we consider the following forms for $f(z)$, $H(z)$, and $h(z)$ up to the first order of $q_2$,
\begin{equation}\label{f}
f(z)=f_0(z)+q_2 f_1(z),
\end{equation}
\begin{equation}\label{H}
H(z)=H_0(z)+q_2 H_1(z)
\end{equation}
\begin{equation}\label{h}
h(z)=h_0(z)+q_2 h_1(z).
\end{equation}
$f_0(z)$ , $h_0(z)$  and $H_0(z)$  is easily found as,
\begin{equation}
f_0(z)= \frac{2M}{z}-\frac{1}{4z}\int^{z}\Big( q_1 h_0'^2 u^2+4 \Lambda  u^2\Big) du=\frac{2 M}{z}+\frac{z^2}{L^2}-\frac{Q^2 q_1 }{8 z^2},
\end{equation}
where,
\begin{equation}\label{m}
M=\frac{ q_1 Q^2}{4 z_h}-\frac{z_h^3}{2L^2}.
\end{equation}
The behavior of $f_0(z)$ in the boundary of AdS, $z \to \infty$, behaves as  $f_0(z) \sim  \frac{z^2}{L^2}$ so the metric on the boundary is as,
\begin{equation}
ds^{2} = \frac{z^2}{L^2}(-dt^{2}+dx^2+dy^2).
\end{equation}
It means that the boundary of AdS is flat and we have $D2$-brane on the boundary.\\
The leading order of gauge field as,
\begin{equation}
h_0(z)=C_1\int_{z_h}^z\frac{du}{3 q_1 u^2}=Q(\frac{1}{z}-\frac{1}{z_h}),
\end{equation}
where $C_1=-3 q_1 Q$,
\begin{equation}
H_0(z)=0.
\end{equation}
We find $H_1(z)$ as,
 \begin{equation}
H_1(z)= \int_{z_h} ^{z}\frac{2  D(u)}{3 u f_0(u)}du,
\end{equation}
where,
\begin{eqnarray}\label{H1}
D(u)=u^2 h_0' \left(f_0'' h_0'-f_0' h_0''\right)+f_0\left(u^2 h_0''^2+u h_0'
\left(5 h_0''+u h_0^{(3)}(u)\right)+h_0'^2\right).
\end{eqnarray}
By inserting $f_0$ , $h_0$, and $H_0$ in $H_1$ eq.(\ref{H1}) we have,
\begin{equation}
H_1(z) = \int_{z_h}^{z}\frac{2 L^2 Q^4 q_1 (u-3 z_h)-8 Q^2 z_h u \left(z_h^3-7 u^3\right)}{3 u^5 \left(-4 u
	z_h^4+4 u^4 z_h+L^2 Q^2 q_1 (u-z_h)\right)}du,
\end{equation}
 $H_1(z)$ on the boundary of AdS, $z \to \infty$, behaves as $H_1(z) \sim \frac{1}{z^4}$.
\begin{equation}
h_1(z)=C_2\int_{z_h}^z \frac{  u \left(-8 f_0'(u)+4 u f_0''(u)+3 q_1 u H_1(u)\right)+8 f_0(u)}{9q_1^2 u^4} du.
\end{equation}
 $h_1(z)$ on the boundary of AdS, $z \to \infty$, behaves as $h_1(z) \sim \ln z$. 
\begin{equation}
f_1(z)=\frac{1}{6z} \int_{z_h}^{z}  E(u)du
\end{equation}
\begin{eqnarray}
&&E(u)=u h_0' \left(-u \left(4 f_0'' h_0'+3 q_1 h_1'\right)+4 f_0' \left(3 h_0'+u h_0''\right)-3 q_1 u H_1 h_0'\right)\nonumber \\ 
&&+8 f_0 \left(u^2 h_0''^2+u h_0' \left(5
h_0''+u h_0^{(3)}(u)\right)+h_0'^2\right)
\end{eqnarray}
 $f_1(z)$ on the boundary of AdS, $z \to \infty$, behaves as $f_1(z) \sim \frac{1}{z}$.
\section{Holographic aspects }
\label{sec3}
\indent   There is a dictionary in AdS/CFT duality that states a field $\phi$ in gravity is dual to a relevant operator $\mathcal{O}_{\phi}$ on the CFT side. For calculating the color non-abelian DC conductivity in this setup, we should turn on a gauge potential perturbation as $\tilde{A}_x=\tilde{A}_x(z)e^{-i\omega t}$ on the gravity background which is dual to the boundary vector current $\mathop{J}^i$ \cite{Hartnoll:2008vx}.\\

The conductivity is related to the retarded correlation function of currents at zero momentum via the Green-Kubo formula,
\begin{equation} \label{kubo2}
\sigma^{ij}_{ab} (k_{\mu})=\frac{1}{\omega}\int dt \theta(t) e^{i \omega t}<[J^i_a(t, \bold{K}),J^j_b(t, \bold{K})]>=-\mathop{\lim }\limits_{\omega \to 0} \frac{1}{\omega } \Im G^{ij}(k_{\mu}).
\end{equation}
 We use the AdS/CFT dictionary to calculate the retarded correlation function of currents. We insert the gauge potential perturbation into the action and expand the resulting action Eq.(\ref{action}) up to the second order of the perturbed part. Therefore, the resulting action is as follows,
 \begin{eqnarray}\label{action-2}
 S^{(2)}=-\int d^4x \frac{e^{-2 i \omega t}}{z f L^2} \Bigg[-f^2 \left((\partial_z\tilde{A}_x^{(1)})^2+(\partial_z\tilde{A}_x^{(2)})^2+(\partial_z\tilde{A}_x^{(3)})^2\right)\bigg(e^{2 i \omega t} L^2 q_1 z + 4 q_2 f'\bigg) \nonumber\\+
 e^{2 H+2 i \omega t}L^2 q_1 z\bigg( \big(\omega^2+h^2\big) ((\tilde{A}_x^{(1)})^2+(\tilde{A}_x^{(2)})^2)+\omega^2(\tilde{A}_x^{(3)})^2  \bigg)+\nonumber\\
 -4e^{2 H} q_2 \Big((\omega^2-h^2)\left( \tilde{A}_x^{(1)})^2+(\tilde{A}_x^{(2)})^2\right)\Big) +\omega ^2 (\tilde{A}_x^{(3)})^2\Big)(f'-2fH')\Bigg].
 \end{eqnarray}
Varying the action $S^{(2)}$ (\ref{action-2}) with respect to $\tilde{A}_x ^{(1)}$ as,\\
 \begin{align}\label{PerA1}
 f \left(f\tilde{A}_x^{(1)'}\right)'- e^{2 H} \tilde{A}_x^{(1)}\left( h^2+ \omega ^2\right)+\frac{q_2}{2 L^2 z q_1} \Big[ E_0+E_1\Big] =0,     
 \end{align}
 where,
 \begin{align} 
 E_0=8 e^{2 H} \tilde{A}_x^{(1)}e^{-2 i \omega t}(\omega^2- h^2)(f'-2fH'),
 \end{align}
 \begin{equation}
 E_1=\frac{-8 e^{-2 i \omega t} f}{z}\bigg[z \tilde{A}_x^{(1)'} f'^2+ f f' \left(-\tilde{A}_x^{(1)'}+z\tilde{A}_x^{(1)''} \right)+zf \tilde{A}_x^{(1)'} f''\bigg]. 
 \end{equation}
Varying the action $S^{(2)}$ (\ref{action-2}) with respect to $\tilde{A}_x ^{(3)}$ as,\\
 \begin{equation}\label{PerA3}
f   \left(f\tilde{A}_x^{(3)'}\right)'-\frac{2 q_2 e^{-4 i  \omega t}}{L^2 q_1 z^2}E_3=0,
\end{equation}
where,
 \begin{equation}
E_3=f \left(4 z f \tilde{A}_x^{(3)''} f'+\tilde{A}_x^{(3)'} \left(4  z f'^2-4  f f'+4  z f
f''\right)\right)+z \omega ^2 \tilde{A}_x^{(3)} e^{2 H} \left(8  f H'-4  f'\right).
\end{equation}
 By using these relations $f_0\sim4\pi f_0'(z_h)(z-z_h)$ and $f_1\sim4\pi f_1'(z_h)(z-z_h)$, we can determine the solution of Eq.(\ref{PerA1}) and Eq.(\ref{PerA3}) near the event horizon. Since $\tilde{A}_x^{(a)}$ should be suppressed on the event horizon, we will consider $\tilde{A}_x^{(a)}$ as follows,
 \begin{align}
 \tilde{A}_x^{(a)}\sim (z-z_h)^{z_a} \, , \qquad a=1,2,3
 \end{align}
 where,
 \begin{eqnarray}\label{z12}
 &z_1=z_2=\pm i \frac{ \sqrt{ \left(16 \pi  q_2 Te^{-2i  \omega t+H(z_h)}+L^2 q_1 z_h \right)h(z_h)^2+\omega ^2 \left(-16 e^{-2i  \omega t+H(z_h)} \pi  q_2 T+L^2 q_1 z_h \right)}}{4 \pi T \sqrt{L^2 q_1 z_h+16 \pi  q_2 T e^{-2 i  \omega t+H(z_h)}}} \\
 &\label{z3}
 z_3=\pm i \frac{\sqrt{L^2 q_1 z_h \omega ^2 -16 \pi  q_2 T \omega ^2 e^{ H(z_h)-2 i \omega t  }}}{4
 	\sqrt{\pi ^2 L^2 q_1 z_h T^2+16 \pi ^3 q_2 T^3 e^{H(z_h)-2 i \omega t  }}}
 \end{eqnarray}
The Hawking temperature of the black brane $T$ is as follows,
\begin{eqnarray}
 &&T=\frac{1}{2 \pi} \Big[ \frac{1}{\sqrt{g_{rr}}}\frac{d}{dz}\sqrt{-g_{tt}}\Big]\Bigg|_{r=z_h}=\frac{e^{-H(z_h)} f'(z_h)}{4 \pi}\nonumber\\
 &&=\frac{6 z_h^4+\mathit{k} L^2 Q^2 q_1}{8 \pi  L^2 z_h^3}+q_2.
\end{eqnarray} 
 For solving the $\tilde{A}_x^{(a)}$  from event horizon to boundary, we consider the following ansatz, 
 \begin{align}\label{EOMA1}
 \tilde{A}_x^{(1)}=\tilde{A}^{(1)}_{\infty}\Big(\frac{-3f}{\Lambda z^2}\Big)^{z_1}\Big(1+i\omega b_1(z)+\cdots\Big) ,
 \end{align}
 \begin{align}\label{EOMA2}
 \tilde{A}_x^{(2)}=\tilde{A}^{(2)}_{\infty}\Big(\frac{-3f}{\Lambda z^2}\Big)^{z_2}\Big(1+i\omega b_2(z)+\cdots\Big) ,
 \end{align}
 \begin{align}\label{EOMA3}
 \tilde{A}_x^{(3)}=\tilde{A}^{(3)}_{\infty}\Big(\frac{-3f}{\Lambda z^2}\Big)^{z_3}\Big(1+i\omega b_3(z)+\cdots\Big) ,
 \end{align}
 where $\tilde{A}^{(a)}_{\infty}$ is the value of fields in the boundary and $z_i$'s are the minus sign of Eq.(\ref{z12}) and Eq.(\ref{z3}). $b_3(z)$  in Eq.(\ref{EOMA3}) is as follows,
\begin{equation}\label{b3}
b_3(z)=C_3+\int^z\left(\frac{u C_{4}}{f \left(q_1 u L^2+4 q_2 f'\right)}-\frac{u f'-2 f}{u f}\right)du,
\end{equation}
 where $C_3$ and $C_4$ are integration constants.\\ 
% \begin{align}
 %b_3(z)=\int^r \Big(\frac{2}{u}+\frac{\frac{C_{13}}{L^2 q_1}-f_0'}{f_0}\Big)du+q_2\int^r \frac{f_1
 %	\left(f_0'-\frac{C_{13}}{L^2 q_1}\right)-f_0 \left(f_1'+\frac{4 C_{13} f_0'}{L^4 q_1^2 u}\right)}{f_0^2}du.
%  \end{align}
 %\begin{align}\label{C4}
% b_3 \approx & \Bigg(\frac{C_{13}}{ L^2 q_1 f_0'(z_h)}-1+q_2\bigg( \frac{f_1'(z_h)
% 	\left(f_0'(z_h)-\frac{C_{13}}{L^2 q_1}\right)-f_0(z_h) \left(f_1'(z_h)+\frac{4 C_{13} f_0'(z_h)}{L^4 q_1^2 u}\right)}{f_0'(z_h)^2}\bigg)\Bigg) \log(r-z_h)\nonumber\\&+\text{finite terms}.
% \end{align}
 The solution should be regular on the event horizon. Therefore, $C_{4}$ is determined by demanding this condition up to the first order of $q_2$ as follows, 
  \begin{equation}
 C_{4}=L^2 q_1 f'(z_h)-\frac{4q_2}{z_h} f'(z_h)^2
 \end{equation}
 Finally, the two-point functions of the vector current are derived by taking the double variation with respect to the value of the gauge field on the boundary\cite{Policastro2002}.\\

Considering the solution of $\tilde{A}_x^{(3)}$ in Eq.(\ref{action-2}) and variation with respect to $\tilde{A}^{(3)}_{\infty}$, Green's function can be read as,
\begin{align} \label{Green1}
 & G_{xx}^{(33)} (\omega ,\vec{0})=-i \omega \frac{C_4 r e^{H(z_h)} \sqrt{L^2 q_1 z_h-4 q_2 f'(z_h)}}{f'(z_h) \left(4 q_2 f'(z)+L^2 q_1
 	r\right) \sqrt{4 q_2 f'(z_h)+L^2 q_1z_h}}\bigg|_{z \to \infty}.
\end{align}
 The conductivity is as following,
 %\begin{eqnarray}\label{sigma33}
 %\sigma_{xx}^{(33)}=-\mathop{\lim }\limits_{\omega \to 0} \frac{1}{\omega } \Im G^{ij}(k_{\mu}) =1+ \frac{ q_2}{z_h L^2 q_1
 %	f_0'(z_h)} \Bigg(4 f_0(z_h) f_0'(z_h)+L^2 q_1 z_h f_0(z_h) f_1'(z_h)\Bigg),
 %\end{eqnarray} 
 \begin{eqnarray}
 \sigma_{xx}^{(33)}=\frac{C_4 r e^{H(z_h)} \sqrt{L^2 q_1 z_h-4 q_2 f'(z_h)}}{f'(z_h) \left(4 q_2 f'(z)+L^2 q_1
 	r\right) \sqrt{4 q_2 f'(z_h)+L^2 q_1z_h}}\bigg|_{z \to \infty}
  \end{eqnarray}
by substituting $f_0(z)$ , $f_1(z)$  in above equation and using Eq.(\ref{kubo2}) we have, 
%\begin{eqnarray} 
% &&\sigma_{xx}^{(33)} = \frac{C_{13}}{L^2 q_1 f_0'(z_h)}+\nonumber\\&&q_2\frac{C_{13}  \Bigg(f_0'(z_h) \left(L^2 q_1 r z_h
% 	H_1(z_h)-4 \left(z_h f_0'(z)+r f_0'(z_h)\right)\right)-L^2 q_1 r z_h
% 	f_1'(z_h)\Bigg)}{L^4 q_1^2 r z_h f_0'(z_h)^2}
%\end{eqnarray} 
\begin{eqnarray}
 &&\sigma_{xx}^{(33)} =1+q_2\Bigg( H_1(z_h)-\frac{4(z_h f_0'(z) +2 r f_0'(z_h) )}{ L^2 r q_1 z_h} \Bigg)\bigg|_{z \to \infty}\nonumber\\&&=1-q_2\Big(\frac{8 f_0'(z_h)}{L^2  q_1 z_h}+\frac{8}{L^4 q_1} -H_1(z_h) \Big)=1-\frac{8q_2}{L^2  q_1 z_h}\bigg(\frac{4z_h}{L^2}+\frac{Q^2 q_1}{2z_h^3}\bigg)
\end{eqnarray}
 It shows that the conductivity bound is violated for the non-minimal $R_{\mu \alpha \nu \beta } F^{\mu \alpha } F^{\nu \beta }$ black brane theory. In the limit of $q_2 \to 0$,  our result recovers the DC conductivity of Einstein-Yang-Mills AdS black brane\cite{Parvizi:2020ses} ,$\sigma_{xx}^{(33)} =1$.\\\\
It can be easily shown that $\sigma_{xx}^{(11)}=\sigma_{xx}^{(22)} =0$.\\
Now, we want to calculate the ratio of shear viscosity to entropy density quantity by using the formula introduced in the \cite{Hartnoll:2016tri}.
\begin{equation}
\frac{\eta}{s}=\frac{1}{4 \pi} \phi(z_h)^2.
\end{equation}
 We perturb the metric as $g_{\mu \nu} \to g_{\mu \nu} +\frac{r^2}{L^2} \phi(r)$ and put it into the action and expand the action up to the second order of $\phi$. Therefore, we derive the equation of motion of $\phi(r)$ as the following by varying the resulting action with respect to $\phi(z)$,
\begin{eqnarray}
\phi ' \Big(r f'+f \left(4-2 r H'\right)\Big)+r f \phi ''=0.
\end{eqnarray}
The solution of the $\phi(z)$ is as follows,
\begin{align}\label{phi}
\phi (z)=C_5 + C_6\int^{r}\frac{e^{2 H(u)} }{u^4 f(u)}du.
\end{align}
We find the solution up to first order of $q_2$ as,
\begin{align}
\phi (z)=\phi_0 (z)+q_2 \phi_1 (z)=C_5+ C_6\int ^{r}\frac{du}{u^4 f_0(u) }+C_6 q_2  \int^r\frac{2 f_0 H_1-f_1}{u^4 f_0^2}du.
\end{align}
The solution of $\phi(z)$ near the event horizon is as the following,
\begin{eqnarray}
&&\phi(z)=\phi_0 (z)+q_2\phi_1 (z)= C_5+  \frac{C_6}{4 \pi T f_0'(z_h) z_h^4}\log(z-z_h)\nonumber\\&&+C_6 q_2 \Bigg(\frac{2 H_1(z_h)  }{z_h^4 f_0'(z_h)}-\frac{f_1'(z_h)  }{z_h^4 f_0'(z_h)^2}\Bigg)\log(z-z_h),
\end{eqnarray}
$\phi (z)$ should be regular on the event horizon $z_h$. By applying this condition we have, 
\begin{eqnarray}
C_6=0
\end{eqnarray}
We set $C_5=1$ for normalization. Then, the solution of $\phi(z)$ is as follows,
\begin{align}
\phi(z)=\phi_0(z)+q_2 \phi_1(z)=1.
\end{align}
Finally, the ratio of shear viscosity to entropy density up to the first order of $q_2$ is as,
\begin{equation}
\frac{\eta}{s}=\frac{1}{4 \pi} \phi(z_h)^2=\frac{1}{4 \pi} (1+2 q_2 \phi_1(z_h) )=\frac{1}{4 \pi}.
\end{equation}
Kovtun, Son, and Starinets (KSS) bound \cite{Policastro2002} states that $\frac{\eta}{s} \geq \frac{1}{4 \pi} $ for all quantum field theories. This bound is saturated for Einstein-Hilbert gravity with a field theory dual and it is violated for higher derivative gravities\cite{Sadeghi:2015vaa,Sadeghi:2022kgi}, degenerate-higher-order-scalar-tensor theories\cite{Bravo-Gaete:2022lno} and Horndeski theory\cite{Bravo-Gaete:2021hlc} and generalized scalar tensor theories\cite{Bravo-Gaete:2020lzs}, massive gravity \cite{Hartnoll:2016tri}\cite{Sadeghi:2015vaa},\cite{Parvizi:2017boc},\cite{Alberte:2016xja} and deformed black brane\cite{Ferreira-Martins:2019svk}. The KSS bound is saturated for our model up to the first order of $q_2$.
%--------------------------------------------------------------------------

 \section{Conclusion}

\noindent We considered Einstein-Yang-Mills theory with a non-minimal coupling between the Riemann and Yang-Mills fields in Anti-de Sitter (AdS) spacetime. The Einstein's equations of this model do not have an exact analytic solution. Therefore, we solved it up to the first order of $q_2$ and introduced the black brane solution for this model. There is a well-known conductivity bound in the literature for Einstein-Hilbert gravity with a field theory dual, which is $\sigma \geq 1$ {\footnote{ We consider $\frac{1}{e^2}=1$}. Our results indicate that this bound is violated for $q_2<0$ in this model. It means that the quantum field theory dual of this model differs from the quantum field theory dual of Einstein-Hilbert gravity. The KSS bound is saturated for this model up to the first order of non-minimal coupling $q_2$. Since $\frac{\eta}{s}$ is proportional to the inverse square of the field theory side coupling, it implies that the coupling of the field theory dual in our model is the same as the coupling of the field theory dual in Einstein-Hilbert gravity.

%--------------------------------------------------------------------------
\vspace{1cm}
\noindent {\large {\bf Acknowledgment} } Author would like to thank Shahrokh Parvizi, Ahmadreza Moradpouri and Komeil Babaei for useful comments and suggestions. Author also thanks to the referees of CQG for the constructive comments and recommendations which definitely help to improve the manuscript.\\

%--------------------------------------------------------------------------
\vspace{1cm}
\noindent {\large {\bf Data Availability} } All data that support the findings of this study are included within the article.
%--------------------------------------------------------------------------

\end{document}